\documentclass[aip,reprint]{revtex4-1}
\usepackage{graphicx}
\usepackage{appendix}
\usepackage{braket}
\usepackage{rotating}
\usepackage{simplewick}
\usepackage{amssymb}
\usepackage{amsmath}
\usepackage{txfonts}
\usepackage{bm}
\usepackage{color}
\usepackage{multirow}
\usepackage{booktabs}
\usepackage{dcolumn}

\begin{document}

\title{Finite-temperature many-body perturbation theory in the grand canonical ensemble}

\author{So Hirata}
\email{sohirata@illinois.edu.}
\author{Punit K. Jha}
\affiliation{Department of Chemistry, University of Illinois at Urbana-Champaign, Urbana, Illinois 61801, USA}

\date{\today}

\begin{abstract}
A finite-temperature many-body perturbation theory is presented that expands in power series the electronic grand potential, chemical potential, internal energy,
and entropy on an equal footing. Sum-over-states and sum-over-orbitals analytical formulas for the second-order perturbation corrections to 
these thermodynamic properties are obtained in a time-independent, nondiagrammatic, algebraic derivation, relying on the sum rules of the Hirschfelder--Certain
degenerate perturbation energies in a degenerate subspace as well as nine algebraic identities for zeroth-order thermal averages of one- through 
four-indexed quantities and products thereof. 
They reproduce numerically exactly the benchmark data obtained as the numerical derivatives of the 
thermal-full-configuration-interaction results for a wide range of temperature.
\end{abstract}

\maketitle 

\section{Introduction}

It is of interest to establish a finite-temperature many-body perturbation theory for electrons 
that expands in power series the 
grand potential ($\Omega$), chemical potential ($\mu$), internal energy ($U$), and entropy ($S$) on an equal footing. 

Recently,\cite{HirataJha}
we derived sum-over-states and reduced analytical formulas for the first-order corrections to these quantities of such a perturbation theory.
Our formulas are based on, and thus agree numerically exactly with,\cite{JhaHirata} 
the postulate of a canonical perturbation theory,\cite{Hirata2017} defining the $n$th-order correction of quantity $X$ by
\begin{eqnarray}
X^{(n)} = \left.\frac{1}{n!} \frac{\partial^n X(\lambda)}{\partial \lambda^n}\right|_{\lambda=0},  \label{lambda}
\end{eqnarray}
where $X(\lambda)$ is evaluated exactly [i.e., by thermal full configuration interaction (FCI)\cite{Kou} in this context] 
with a scaled Hamiltonian
$\hat{H} = \hat{H}_0 + \lambda \hat{V}$. This is fully equivalent to the usual perturbation expansion of $X$:
\begin{eqnarray}
X &=& X^{(0)} + \lambda X^{(1)} + \lambda^2 X^{(2)} + \lambda^3 X^{(3)} + \dots.
\end{eqnarray}
This differs from the earlier definition\cite{bloch,kohn,luttingerward,balian,blochbook,Zgid,Zgid2,SANTRA} of a finite-temperature many-body perturbation theory for electrons, 
described in many textbooks,\cite{thouless1972quantum,mattuck1992guide,march1995many,Fetter} 
which variationally determines $\mu$ at each order 
to restore the electrical neutrality.\cite{JhaHirata,Levin,Fisher,Dyson,HirataARPC} 
Such a series is converging at thermal FCI, but may not agree with Eq.\ (\ref{lambda}).

In this article, we present sum-over-states and reduced (sum-over-orbitals) analytical formulas for the second-order corrections to $\Omega$, $\mu$, and $U$ 
(from which the same for $S$ can be inferred) derived in a time-independent, algebraic (nondiagrammatic)
manner. Our derivation is transparent and general, relying only on 
the Hirschfelder--Certain degenerate perturbation theory\cite{Hirschfelder} 
and elementary combinatorics and calculus. This introduces a new derivation strategy in lieu of the usual quantum-field-theoretical one 
(time-dependent diagrammatic techniques using 
Matsubara Green's function\cite{matsubara} or thermal Wick's theorem\cite{sanyal}) whose applicability to $\mu$ and $U$ is yet to be established.

\section{Theory}

Here, we are exclusively concerned with thermodynamics of electrons in the grand canonical ensemble. 
For the sake of completeness and because higher-order perturbation corrections are given in terms of lower-order corrections,
we reiterate the derivations of the zeroth-\cite{thouless1972quantum,mattuck1992guide,march1995many,Fetter} and first-order\cite{HirataJha} perturbation theory formulas as succinctly as possible. For the electronic thermodynamics in the canonical ensemble, see Ref.\ \onlinecite{JhaHirata_canonical}.

The grand partition function $\Xi$ is defined by
\begin{eqnarray}
\Xi = \sum_I e^{- \beta E_I + \beta \mu N_I} , \label{Xidef}
\end{eqnarray}
where $\beta = (k_\text{B}T)^{-1}$ is the inverse temperature, $E_I$ is the exact (zero-temperature FCI) energy of the $I$th state, and $N_I$ is the number of electrons 
in the same state. In the following capital-letter indices $I$ and $J$ run over all $2^n$ states (where $n$ is the number of spinorbitals),
while lower-letter indices refer to spinorbitals. The exact chemical potential $\mu$ is determined by solving the equation,
\begin{eqnarray}
\bar{N} &=& \frac{1}{\beta} \frac{\partial}{\partial \mu}\ln \Xi \label{Nbar0} \\
&=& \frac{ \sum_I N_I e^{- \beta E_I + \beta \mu N_I} }{\sum_I e^{- \beta E_I + \beta \mu N_I}  }, \label{Nbar}
\end{eqnarray}
or
\begin{eqnarray}
\bar{N}{\sum_I e^{- \beta E_I + \beta \mu N_I}  }  = { \sum_I N_I e^{- \beta E_I + \beta \mu N_I} }, \label{Nbar2}
\end{eqnarray}
where $\bar{N}$ is the correct average number of electrons that keeps the system electrically neutral.  

Once $\Xi$ is determined, we can extract from it various thermodynamic properties such as the exact $\Omega$, $U$, and $S$:
\begin{eqnarray}
\Omega &=& -\frac{1}{\beta} \ln \Xi, \label{Omegadef}\\
U &=& - \frac{\partial}{\partial \beta} \ln \Xi + \mu \bar{N} \label{Udef1} \\
&=& \frac{ \sum_I E_I  e^{- \beta E_I + \beta \mu N_I} }{\sum_I e^{- \beta E_I + \beta \mu N_I} }, \label{Udef2} \\
S &=& T^{-1} \left( U -  \mu \bar{N} -  \Omega\right) .
\end{eqnarray}
By ``exact,'' we mean that they are determined by thermal FCI (Ref.\ \onlinecite{Kou}) within a basis set. 
In this procedure, it is Eqs.\ (\ref{Nbar}) and (\ref{Udef2}) [not Eqs.\ (\ref{Nbar0}) or (\ref{Udef1})] that we actually use when numerically determining $\mu$ and $U$
(in other words, nowhere in our thermal FCI program is  numerical differentiation with respect to $\mu$ or $\beta$ performed).
Since $S$ is readily inferred from $\Omega$, $\mu$, and $U$, we will not discuss it any further.

In our finite-temperature many-body perturbation theory, all thermodynamic properties are expanded in perturbation series on an equal footing:
\begin{eqnarray}
\mu &=& \mu^{(0)} + \lambda \mu^{(1)} + \lambda^2 \mu^{(2)} + \lambda^3 \mu^{(3)} + \dots, \\
\Omega &=& \Omega^{(0)} + \lambda \Omega^{(1)} + \lambda^2 \Omega^{(2)} + \lambda^3 \Omega^{(3)} + \dots, \\
U &=& U^{(0)} + \lambda U^{(1)} + \lambda^2 U^{(2)} + \lambda^3 U^{(3)} + \dots,
\end{eqnarray}
which are equivalent to Eq.\ (\ref{lambda}) with $X=\mu$, $\Omega$, or $U$, respectively, where 
$\lambda$ is the dimensionless perturbation strength in the partitioned Hamiltonian:
\begin{eqnarray}
\hat{H} &=& \hat{H}_0 + \lambda \hat{V}. \label{partitioning0}
\end{eqnarray}

In what follows, we adopt the so-called M{\o}ller--Plesset partitioning of the Hamiltonian,
\begin{eqnarray}
\hat{H}_0 &=& E_{\text{nuc.}} + \sum_p \epsilon_p \hat{p}^\dagger \hat{p}, \label{partitioning}  
\end{eqnarray}
where $E_{\text{nuc.}}$ is the nuclear-repulsion energy, $\epsilon_p$ is the canonical Hartree--Fock (HF) energy 
of the $p$th spinorbital, and $\hat{p}^\dagger$ ($\hat{p}$) are the creation (annihilation) operator of an electron in the $p$th spinorbital.
With this choice of $\hat{H}_0$, the perturbation expansions of the $I$th-state energy,
\begin{eqnarray}
E_I &=& E_I^{(0)} + \lambda E_I^{(1)} + \lambda^2 E_I^{(2)} + \lambda^3 E_I^{(3)} + \dots, 
\end{eqnarray}
are given by the Hirschfelder--Certain degenerate perturbation theory (HCPT),\cite{Hirschfelder} which reduces 
to the M{\o}ller--Plesset perturbation theory (MPPT)\cite{szabo,shavitt} for nondegenerate states. Our theory is not fundamentally limited to this
partitioning and can be adjusted (if not directly used) for other cases (such as a finite-temperature HF reference).

The following Taylor expansions are used in the subsequent sections:
\begin{eqnarray}
e^{a+b} &=& e^a + e^a b + e^a \frac{b^2}{2!} + e^a \frac{b^3}{3!} + \dots, \label{exponential} \\
\ln (a+b) &=& \ln a + \frac{b}{a} - \frac{b^2}{2 a^2} + \frac{b^3}{3a^3} + \dots, \label{logarithm}
\end{eqnarray}
where $a \gg b$. 

Our derivation benefitted from an excellent review by Santra and Schirmer\cite{SANTRA} of the conventional
finite-temperature many-body perturbation theory\cite{bloch,kohn,luttingerward,balian,blochbook}
described in many textbooks,\cite{thouless1972quantum,mattuck1992guide,march1995many,Fetter} which differs from the one presented here 
at the first order and higher. 

\section{Zeroth order\cite{thouless1972quantum,mattuck1992guide,march1995many,Fetter}}

The zeroth-order theory is the Fermi--Dirac theory.\cite{thouless1972quantum,mattuck1992guide,march1995many,Fetter}
The zeroth-order grand partition function is
\begin{eqnarray}
\Xi^{(0)} &=& \sum_I e^{- \beta E^{(0)}_I + \beta \mu^{(0)} N_I} ,
\end{eqnarray}
with
\begin{eqnarray}
E_I^{(0)} &=& E_\text{nuc.} + \sum_i^I \epsilon_i, \\
N_I &=& \sum_i^I 1,
\end{eqnarray}
where $\sum_i^I$ means that $i$ runs over all spinorbitals that are occupied in the $I$th Slater-determinant state. 
Henceforth, we use letters $i$, $j$, and $k$ for spinorbitals occupied in the $I$th determinant, $a$, $b$, and $c$ for those unoccupied 
in the same determinant, and $p$, $q$, $r$, and $s$ for either.  

\subsection{Grand potential}

Substituting these into Eq.\ (\ref{Omegadef}), we obtain a sum-over-states formula for $\Omega^{(0)}$ as
\begin{eqnarray}
\Omega^{(0)} &=& -\frac{1}{\beta} \ln \Xi^{(0)} \\
&=& -\frac{1}{\beta} \ln \sum_I e^{- \beta E_I^{(0)} + \beta \mu^{(0)} N_I} \label{Omega0_SoS}\\
&=& E_\text{nuc.} -\frac{1}{\beta} \ln \sum_I e^{\sum_i^I \nu_i},
\end{eqnarray}
with $\nu_i = - \beta (\epsilon_i - \mu^{(0)})$, where $\mu^{(0)}$ will be discussed in the next subsection. 

Using the identity, 
\begin{eqnarray}
\sum_I e^{\sum_i^I \nu_i} &=& \sum_{I_0} 1 + \sum_{I_1} \sum_i^{I_1} e^{\nu_i } + \sum_{I_2} \sum_{i<j}^{I_2} e^{\nu_i}e^{\nu_j} + \dots \nonumber\\
&=& \prod_p \left(1+e^{\nu_p} \right)= \prod_p \frac{ 1 + e^{-\nu_p} }{e^{-\nu_p}} = \prod_p \frac{1}{f_p^+}, \label{identity0}
\end{eqnarray}
where $I_m$ stands for a Slater determinant with $m$ electrons ($0 \leq m \leq n$)
and $f_p^+$ ($f_p^-$) is the Fermi--Dirac vacancy (occupancy) given by
\begin{eqnarray}
f_p^- &=& \frac{1}{1+e^{\beta(\epsilon_p - \mu^{(0)})}}, \\
f_p^+ &=& 1-f_p^- = \frac{e^{\beta(\epsilon_p - \mu^{(0)})}}{1+e^{\beta(\epsilon_p - \mu^{(0)})}}, 
\end{eqnarray}
we obtain a sum-over-orbitals (`reduced') formula for $\Omega^{(0)}$ as
\begin{eqnarray}
\Omega^{(0)} &=& E_\text{nuc.} + \frac{1}{\beta} \sum_p \ln f_p^+ , \label{Omega0_reduced}
\end{eqnarray}
where $p$ runs over all spinorbitals. While the sum-over-states formula [Eq.\ (\ref{Omega0_SoS})] involves
an exponentially long ($2^n$) summation, the reduced formula [Eq.\ (\ref{Omega0_reduced})] accumulates only $n$ terms.

\subsection{Chemical potential}

The sum-over-states equation to be solved for $\mu^{(0)}$ is
\begin{eqnarray}
\bar{N} &=& \frac{ \sum_I N_I e^{- \beta E^{(0)}_I + \beta \mu^{(0)} N_I} }{\sum_I e^{- \beta E^{(0)}_I + \beta \mu^{(0)} N_I}  } \equiv \langle N_I \rangle, \label{mu0_SoS}
\end{eqnarray}
where we introduced a shorthand notation of a zeroth-order thermal average:
\begin{eqnarray}
\langle X_I \rangle \equiv   \frac{ \sum_I X_I e^{- \beta E^{(0)}_I + \beta \mu^{(0)} N_I} }{\sum_I e^{- \beta E^{(0)}_I + \beta \mu^{(0)} N_I}  }. \label{average}
\end{eqnarray}
It can be reduced to a sum-over-orbitals expression by rewriting the right-hand side of Eq.\ (\ref{mu0_SoS}) as
\begin{eqnarray}
\bar{N} &=& \frac{ \sum_{I_1} \sum_i^{I_1} a_i e^{\nu_i } + \sum_{I_2}\sum_{i<j}^{I_2} (a_i + a_j)e^{\nu_i}e^{\nu_j} + \dots }{\prod_p (1+ e^{\nu_p})} \\
&=& \sum_p f_p^-, \label{mu0_reduced}
\end{eqnarray}
where $a_i = 1$, $\nu_i = -\beta (\epsilon_i - \mu^{(0)})$ and the common factor of $e^{-\beta E_{\text{nuc.}}}$ has been canceled between the numerator and denominator.
The second equality follows\cite{HirataJha} immediately from Boltzmann-sum identity I of Appendix \ref{BoltzmannIdentities}.

Alternatively, comparing the sum-over-states formulas of $\Omega^{(0)}$ [Eq.\ (\ref{Omega0_SoS})] and $\mu^{(0)}$ [Eq.\ (\ref{mu0_SoS})], 
we find\cite{thouless1972quantum,mattuck1992guide,march1995many,Fetter}
\begin{eqnarray}
\bar{N} &=&  \frac{1}{\beta} \frac{\partial}{\partial \mu^{(0)}} \ln \Xi^{(0)} \\
&=& -\frac{\partial \Omega^{(0)}}{\partial \mu^{(0)}} = \sum_p f_p^-,
\end{eqnarray}
where the last equality follows by substituting the reduced formula of $\Omega^{(0)}$ [Eq.\ (\ref{Omega0_reduced})] and using
\begin{eqnarray}
\frac{\partial f_p^\pm}{\partial \mu^{(0)}} = \mp \beta f_p^- f_p^+ . \label{mu_derivative}
\end{eqnarray}

\subsection{Internal energy\label{sec:U0}}

The sum-over-states formula for $U^{(0)}$ reads
\begin{eqnarray}
U^{(0)} &=& \frac{ \sum_I E^{(0)}_I e^{- \beta E^{(0)}_I + \beta \mu^{(0)} N_I} }{\sum_I e^{- \beta E^{(0)}_I + \beta \mu^{(0)} N_I}  } \equiv \langle E_I^{(0)} \rangle . \label{U0_SoS}
\end{eqnarray}
Setting $a_i = \epsilon_i$ and $\nu_i = -\beta (\epsilon_i - \mu^{(0)})$ and using Boltzmann-sum identity I of Appendix \ref{BoltzmannIdentities},\cite{HirataJha} we obtain the reduced formula,
\begin{eqnarray}
U^{(0)} &=& E_{\text{nuc.}} + \frac{ \sum_{I_1} \sum_i^{I_1} a_i e^{\nu_i } + \sum_{I_2}\sum_{i<j}^{I_2} (a_i + a_j)e^{\nu_i}e^{\nu_j} + \dots }{\prod_p (1+ e^{\nu_p})}  \nonumber\\
&=& E_{\text{nuc.}} + \sum_p  \epsilon_p f_p^-.  \label{U0_reduced}
\end{eqnarray}

Comparing the sum-over-states formulas of $\Omega^{(0)}$ [Eq.\ (\ref{Omega0_SoS})] and $U^{(0)}$ [Eq.\ (\ref{U0_SoS})], we notice that they are related by
\begin{eqnarray}
U^{(0)} &=& -\frac{\partial \ln \Xi^{(0)}}{\partial \beta} + \mu^{(0)} \bar{N} + \beta \frac{\partial \mu^{(0)} }{\partial \beta}\bar{N} \\
&=& \Omega^{(0)}  + \mu^{(0)} \bar{N}  + \beta \left( \frac{\partial \Omega^{(0)} }{\partial \beta}\right)_{\mu^{(0)}},
\end{eqnarray}
where the subscript $\mu^{(0)}$ indicates that it is held fixed when the partial derivative with respect to $\beta$ is taken. While $f_{p}^\pm$ and $\mu^{(0)}$
vary with $\beta$, the $\beta$-derivative must precede the $\lambda$-derivative (or the perturbation expansion), and, therefore, $\partial \mu^{(0)} / \partial \beta$
should not be taken. Substituting the reduced formula of $\Omega^{(0)}$ [Eq.\ (\ref{Omega0_reduced})] into the above as well as using
\begin{eqnarray}
\frac{\partial f_p^\pm}{\partial \beta} = \pm \left( \epsilon_p - \mu^{(0)} \right)  f_p^- f_p^+ , \label{beta_derivative}
\end{eqnarray}
we arrive at the 
same reduced formula for $U^{(0)}$ given by Eq.\ (\ref{U0_reduced}).

\section{First order\cite{HirataJha}}

Using the Taylor expansion of an exponential [Eq.\ (\ref{exponential})], we obtain 
the sum-over-states formula for $\Xi^{(1)}$, which reads 
\begin{eqnarray}
\Xi^{(1)} = \sum_I \left( - \beta E^{(1)}_I + \beta \mu^{(1)} N_I \right) e^{- \beta E^{(0)}_I + \beta \mu^{(0)} N_I}.
\end{eqnarray}

\subsection{Grand potential}

Expanding the logarithm [Eq.\ (\ref{logarithm})] in the definition of $\Omega$ [Eq.\ (\ref{Omegadef})], we obtain the sum-over-states formula of $\Omega^{(1)}$ as,
\begin{eqnarray}
\Omega^{(1)} &=& -\frac{1}{\beta} \frac{\Xi^{(1)}}{\Xi^{(0)}}  \label{Omega1_2} \\
&=& \langle E_I^{(1)}  - \mu^{(1)} N_I \rangle = \langle E_I^{(1)} \rangle - \mu^{(1)} \bar{N}, \label{Omega1_SoS}
\end{eqnarray}
where 
$\langle N_I \rangle = \bar{N}$ according to Eq.\ (\ref{mu0_SoS}).

At first glance, reducing $\langle E_I^{(1)} \rangle$ into a sum-over-orbitals formula appears implausible because 
there is no closed (or diagrammatic) expression for $E_I^{(1)}$ given in terms of molecular integrals when its zeroth-order energy $E_I^{(0)}$ is degenerate. 
However, the sum of all $E_I^{(1)}$ in a degenerate subspace does have a closed formula,\cite{HirataJha}
i.e., Eq.\ (\ref{E1sumrule}) of Appendix \ref{Sumrules}.
Because these $E_I^{(1)}$ in the degenerate subspace 
are summed with an equal weight of $e^{-\beta E_I^{(0)} + \beta \mu^{(0)} N_I }$, we can in fact reduce $\langle E_I^{(1)} \rangle$
by using this sum rule without knowing individual $E_I^{(1)}$ for each state.

For the purpose of simplifying $\langle E_I^{(1)} \rangle$, therefore, we can pretend that
\begin{eqnarray}
E_I^{(1)} = \sum_i^I a_i + \sum_{i<j}^I b_{ij},
\end{eqnarray}
is true for each state, with $a_i = H_{ii}^{\text{core}} - \epsilon_i $ and $b_{ij} = \langle ij || ij \rangle$, where
$\bm{H}^{\text{core}}$ is the one-electron part of the Fock matrix and $\langle pq||rs \rangle$
is an antisymmetrized two-electron integral. Combining these with Boltzmann-sum
identities I and III of Appendix \ref{BoltzmannIdentities}, we obtain
\begin{widetext}
\begin{eqnarray}
\langle E_I^{(1)} \rangle &=&  \frac{  \sum_{I_1} \sum_i^{I_1} a_i e^{\nu_i } + \sum_{I_2} \sum_{i<j}^{I_2} (a_i + a_j)e^{\nu_i}e^{\nu_j} + \dots  }{\prod_p (1+ e^{\nu_p})} 
+ \frac{  \sum_{I_2} \sum_{i<j}^{I_2} b_{ij} e^{\nu_i}e^{\nu_j} +  \sum_{I_3}\sum_{i<j<k}^{I_3} (b_{ij}+b_{ik}+b_{jk}) e^{\nu_i}e^{\nu_j}e^{\nu_k} + \dots  }{\prod_p (1+ e^{\nu_p})} \\
&=&\sum_p \left( H_{pp}^{\text{core}} - \epsilon_p \right) f_p^- +  \frac{1}{2} \sum_{p,q} \langle pq || pq \rangle f_p^- f_q^-   = \sum_p F_{pp} f_p^- -\frac{1}{2} \sum_{p,q} \langle pq || pq \rangle f_p^- f_q^-, \label{E1}
\end{eqnarray}
\end{widetext}
where $\bm{F}$ is the finite-temperature Fock matrix\cite{SANTRA} minus the diagonal zero-temperature Fock matrix:\cite{szabo}
\begin{eqnarray}
F_{pq} = H_{pq}^{\text{core}} + \sum_r \langle pr || qr \rangle f_r^- -\delta_{pq} \epsilon_p, \label{Fock}
\end{eqnarray}
corresponding to the M{\o}ller--Plesset partitioning [Eq.\ (\ref{partitioning})] we adopt here.

The reduced formula for $\Omega^{(1)}$ is, therefore,
\begin{eqnarray}
\Omega^{(1)} &=& \sum_p F_{pp} f_p^- -\frac{1}{2} \sum_{p,q} \langle pq || pq \rangle f_p^- f_q^- - \mu^{(1)} \bar{N}, \label{Omega1_reduced}
\end{eqnarray}
which differs from the one found in the textbooks\cite{SANTRA,thouless1972quantum,mattuck1992guide,march1995many,Fetter} by the presence of the last term. 
It is the same as Eq.\ (46) of our earlier paper.\cite{HirataJha}

\subsection{Chemical potential}

Expanding the electroneutrality condition [Eq.\ (\ref{Nbar2})] with the Taylor expansion of an exponential [Eq.\ (\ref{exponential})] and 
collecting the first-order terms, we obtain
\begin{eqnarray}
&& \bar{N} \sum_I \left( -\beta E^{(1)}_I + \beta \mu^{(1)} N_I \right) 
e^{- \beta E^{(0)}_I + \beta \mu^{(0)} N_I}  \nonumber \\
&& = \sum_I N_I \left(  -\beta E^{(1)}_I + \beta \mu^{(1)} N_I \right) e^{- \beta E^{(0)}_I + \beta \mu^{(0)} N_I},  \label{1NminusN1}
\end{eqnarray}
or 
\begin{eqnarray}
 \bar{N} \langle  E^{(1)}_I - \mu^{(1)} N_I \rangle 
 = \langle N_I ( E^{(1)}_I - \mu^{(1)} N_I) \rangle.  \label{1NminusN2}
\end{eqnarray}
This can be solved for $\mu^{(1)}$, leading to its sum-over-states formula:
\begin{eqnarray}
\mu^{(1)} = \frac{\langle E_I^{(1)} (N_I - \bar{N})\rangle }{\langle N_I (N_I - \bar{N}) \rangle }
=   \frac{\langle E_I^{(1)} N_I \rangle - \langle E_I^{(1)} \rangle \bar{N} }{\langle N_I^2 \rangle - \bar{N}^2 }. \label{mu1_SoS}
\end{eqnarray}
Using Boltzmann-sum identities II and IV of Appendix \ref{BoltzmannIdentities}, we can simplify the numerator and denominator as\cite{HirataJha}
\begin{eqnarray}
\langle E_I^{(1)} N_I \rangle - \langle E_I^{(1)} \rangle \bar{N} &=& \sum_p F_{pp} f_p^- f_p^+, \label{E1N} \\
\langle N_I^2 \rangle - \bar{N}^2 &=& \sum_p f_p^- f_p^+, \label{NN}
\end{eqnarray}
arriving at the reduced formula for $\mu^{(1)}$ that reads
\begin{eqnarray}
\mu^{(1)} = \frac{\sum_p F_{pp} f_p^- f_p^+ }{\sum_p f_p^- f_p^+}. \label{mu1_reduced}
\end{eqnarray}
This is identified as Eq.\ (48) of our earlier paper.\cite{HirataJha}

Alternatively, differentiating the sum-over-states formula of $\Omega^{(1)}$ [Eq.\ (\ref{Omega1_2})] with respect to $\mu^{(0)}$ (while holding $\mu^{(1)}$ fixed) and using 
the first-order electroneutrality [Eq.\ (\ref{1NminusN2})], we find
\begin{eqnarray}
\left( \frac{\partial \Omega^{(1)}}{\partial \mu^{(0)}} \right)_{\mu^{(1)}} = 0. \label{Omega1mu0}
\end{eqnarray}
Substituting the reduced formula of $\Omega^{(1)}$ [Eq.\ (\ref{Omega1_reduced})] into this and noting $\bar{N} = \sum_p f_p^-$ as well as 
Eq.\ (\ref{mu_derivative}), 
we obtain the same reduced formula for $\mu^{(1)}$ [Eq.\ (\ref{mu1_reduced})].

\subsection{Internal energy}

Expanding an exponential [Eq.\ (\ref{exponential})] in the definition of $U$ [Eq.\ (\ref{Udef2})], we obtain the sum-over-states formula:
\begin{eqnarray}
U^{(1)} 
&=& \langle E_I^{(1)} \rangle - \beta \langle E_I^{(0)}  ( E_I^{(1)} - \mu^{(1)}N_I )  \rangle \nonumber \\
&& + \beta\langle E_I^{(0)} \rangle  \langle E_I^{(1)} - \mu^{(1)}N_I \rangle  \label{U1_SoS}.
\end{eqnarray}
Starting with Eq.\ (\ref{Udef1}) [instead of Eq.\ (\ref{Udef2})] needs some caution. 
As discussed in Sec.\ \ref{sec:U0}, 
since the $\beta$-derivative must precede the perturbation expansion, 
the derivative of $\mu^{(0)}$ or $\mu^{(1)}$ with respect to $\beta$ should not be taken. This fact may be obscured
if Eq.\ (\ref{Udef1}) were used as a starting point. 

Keeping this in mind, we differentiate the sum-over-states formula of $\Omega^{(1)}$ [Eq.\ (\ref{Omega1_2})] and find
\begin{eqnarray}
\frac{\partial}{\partial \beta} \left( \beta \Omega^{(1)} \right) 
&=& -\frac{\partial}{\partial \beta}\left( \frac{\Xi^{(1)}}{\Xi^{(0)}} \right) \\
&=& \langle E_I^{(1)} -  \mu^{(1)}N_I \rangle 
\nonumber\\&& 
- \beta \langle ( E_I^{(1)} - \mu^{(1)}N_I )(E_I^{(0)} -\mu^{(0)}N_I) \rangle 
\nonumber \\&& 
+ \beta \langle E_I^{(1)} - \mu^{(1)}N_I \rangle \langle E_I^{(0)} -\mu^{(0)}N_I \rangle 
\nonumber\\&& 
 +\beta^2 \frac{\partial \mu^{(0)}}{\partial \beta} \langle (E_I^{(1)} -  \mu^{(1)}N_I)N_I \rangle 
\nonumber \\&& 
 - \beta^2 \frac{\partial \mu^{(0)}}{\partial \beta} \langle E_I^{(1)} -  \mu^{(1)}N_I \rangle \langle N_I \rangle 
- \beta \frac{\partial \mu^{(1)}}{\partial \beta} \langle N_I \rangle 
\nonumber \\
&=& \langle E_I^{(1)} \rangle - \beta \langle E_I^{(0)} ( E_I^{(1)} - \mu^{(1)}N_I )\rangle 
\nonumber \\&& 
+ \beta \langle E_I^{(0)} \rangle \langle E_I^{(1)} - \mu^{(1)}N_I \rangle  
-  \mu^{(1)}\bar{N} - \beta \frac{\partial \mu^{(1)}}{\partial \beta} \bar{N} , \nonumber\\ \label{betabetaOmega1}
\end{eqnarray}
where the first-order electroneutrality [Eq.\ (\ref{1NminusN2})] was used twice in the last equality. Comparing the last expression with Eq.\ (\ref{U1_SoS}), 
we notice that $U^{(1)}$ and $\Omega^{(1)}$ are related to each other by
\begin{eqnarray}
U^{(1)} &=& \frac{\partial}{\partial \beta}\left( \beta \Omega^{(1)}\right) + \mu^{(1)} \bar{N} + \beta \frac{\partial \mu^{(1)}}{\partial \beta} \bar{N} \\
&=& \Omega^{(1)} + \mu^{(1)} \bar{N} + \beta \left( \frac{\partial \Omega^{(1)}}{\partial \beta} \right) _{\mu^{(0)},\,\mu^{(1)}} \label{U1fromOmega1} \\
&=& \sum_p F_{pp} f_p^- -\frac{1}{2} \sum_{p,q} \langle pq || pq \rangle f_p^- f_q^- 
\nonumber \\&& 
- \beta \sum_p  \left(  F_{pp} -\mu^{(1)} \right)  \epsilon_p   f_p^- f_p^+, \label{U1_reduced}
\end{eqnarray}
where the reduced formula for $\Omega^{(1)}$ [Eq.\ (\ref{Omega1_reduced})] was substituted in the last equality. Equation (\ref{U1_reduced}) is the reduced formula for $U^{(1)}$
and can be identified as Eq.\ (49) of Ref.\ \onlinecite{HirataJha}.

We can, therefore, start with Eq.\ (\ref{Udef1}) and still obtain a useful relationship between $U^{(1)}$ and $\Omega^{(1)}$,
insofar as care is exercised to ensure that $\mu^{(0)}$ and $\mu^{(1)}$ are held fixed in the partial differentiation with $\beta$. 
Dropping $\mu^{(0)}$ as a fixed variable is permitted because terms involving the derivative of $\mu^{(0)}$ 
cancel with each other. It is, however, incorrect to drop $\mu^{(1)}$ as a fixed variable.\cite{HirataJha}

\section{Second order}

Using the Taylor expansion of an exponential [Eq.\ (\ref{exponential})] to $\Xi$, we find the sum-over-states formula for $\Xi^{(2)}$ as
\begin{eqnarray}
\Xi^{(2)} &=& \sum_I \left\{- \beta E^{(2)}_I + \beta \mu^{(2)} N_I + \frac{1}{2}\left( - \beta E^{(1)}_I + \beta \mu^{(1)} N_I \right)^2 \right\} 
\nonumber\\&& \times\,e^{- \beta E^{(0)}_I + \beta \mu^{(0)} N_I} .
\end{eqnarray}

\subsection{Grand potential}

Likewise, using the Taylor expansion of a logarithm [Eq.\ (\ref{logarithm})], we obtain the sum-over-states formula for $\Omega^{(2)}$ as
\begin{eqnarray}
\Omega^{(2)} &=& -\frac{1}{\beta} \frac{\Xi^{(2)}}{\Xi^{(0)}} + \frac{1}{2\beta} \left(\frac{\Xi^{(1)}}{\Xi^{(0)}} \right)^2 \label{Omega2_start} \\
&=& \langle E_I^{(2)}  - \mu^{(2)} N_I \rangle -\frac{\beta}{2} \langle (E_I^{(1)} - \mu^{(1)}N_I)^2 \rangle \nonumber\\
&& + \frac{\beta}{2} \langle E_I^{(1)}- \mu^{(1)}N_I \rangle^2 .  \label{Omega2_SoS} 
\end{eqnarray}

From this expression, combining the sum rules for $E_I^{(2)}$ and $(E_I^{(1)})^2$ in Appendix \ref{Sumrules} with the nine Boltzmann-sum identities in Appendix \ref{BoltzmannIdentities}, 
after some work (expounded on in Appendix \ref{DerivationOmega2}), we arrive at the following reduced formula:
\begin{eqnarray}
\Omega^{(2)} &=& \sum_{p,q}^{\text{denom.}\neq 0} \frac{| F_{pq} |^2 f_p^- f_q^+}{\epsilon_p-\epsilon_q}
+ \frac{1}{4} \sum_{p,q,r,s}^{\text{denom.}\neq 0} \frac{| \langle pq || rs \rangle |^2 f_p^- f_q^- f_r^+ f_s^+}{\epsilon_p+\epsilon_q-\epsilon_r-\epsilon_s} 
\nonumber\\&& 
- \frac{\beta}{2} \sum_{p,q}^{\text{denom.}= 0} {| F_{pq} |^2 f_p^- f_q^+}
\nonumber\\&& 
- \frac{\beta}{8} \sum_{p,q,r,s}^{\text{denom.}= 0} {| \langle pq || rs \rangle |^2 f_p^- f_q^- f_r^+ f_s^+} 
\nonumber\\&& 
- \mu^{(2)} \bar{N} + \beta \mu^{(1)} \sum_p F_{pp}f_p^- f_p^+ - \frac{\beta}{2} \left(\mu^{(1)}\right)^2 \sum_p f_p^- f_p^+ , \label{Omega2_reduced} 
\end{eqnarray}
where ``$\text{denom.}\neq 0$'' in the first term means that the sums are taken over only $p$ and $q$ that satisfy $\epsilon_p-\epsilon_q\neq 0$, 
while the same in the second term demands $\epsilon_p+\epsilon_q-\epsilon_r-\epsilon_s \neq 0$.
In the third and fourth terms, the sums are taken over cases where 
$\epsilon_p-\epsilon_q=0$ or $\epsilon_p+\epsilon_q-\epsilon_r-\epsilon_s=0$, respectively.
The penultimate  term is identified as $(-2)$ times the last term and, therefore, they can be consolidated. However, we leave them separate 
to keep track of the origin of each factor for later use [in Eqs.\ (\ref{Omega2mu0}) and (\ref{U2fromOmega2}), specifically].

This reduced formula differs from the one in textbooks\cite{SANTRA,thouless1972quantum,mattuck1992guide,march1995many,Fetter} in
the last three terms all involving perturbation corrections to $\mu$. The first of these ($-\mu^{(2)}\bar{N}$) is expected, but the other two
may not be.

The derivation of this formula is tedious but straightforward, using the identical strategy as 
that leading to $\Omega^{(1)}$. It is given in detail in Appendix \ref{DerivationOmega2}.

\subsection{Chemical potential}

Collecting the second-order terms in the electroneutrality condition [Eq.\ (\ref{Nbar2})], we obtain
\begin{eqnarray}
&& 
\bar{N} \left(  \langle E^{(2)}_I -  \mu^{(2)} N_I \rangle - \frac{\beta}{2}  \langle (E^{(1)}_I - \mu^{(1)} N_I)^2 \rangle  \right) = 
\nonumber \\&& 
\langle N_I (E^{(2)}_I -  \mu^{(2)} N_I) \rangle - \frac{\beta}{2}  \langle N_I ( E^{(1)}_I - \mu^{(1)} N_I )^2 \rangle , \label{2NminusN1}
\end{eqnarray}
which can be solved for $\mu^{(2)}$ to give the sum-over-states formula:
\begin{eqnarray}
\mu^{(2)} &=& \frac{\langle E^{(2)}_I  (N_I - \bar{N})\rangle }{\langle N_I(N_I - \bar{N}) \rangle} 
\nonumber\\&&  
- \frac{\beta}{2} \frac{\langle ( E^{(1)}_I - \mu^{(1)} N_I )^2 (N_I - \bar{N})\rangle }{\langle N_I(N_I - \bar{N}) \rangle} . \label{mu2_SoS} 
\end{eqnarray}

Differentiating the sum-over-states formula of $\Omega^{(2)}$ [Eq.\ (\ref{Omega2_start})] with respect to $\mu^{(0)}$ (while holding $\mu^{(1)}$ and $\mu^{(2)}$ fixed) 
and using Eqs.\ (\ref{Omega1mu0}) and (\ref{2NminusN1}), we find
\begin{eqnarray}
\left(\frac{\partial \Omega^{(2)}}{\partial \mu^{(0)}}\right)_{\mu^{(1)},\,\mu^{(2)}}  = 0 . \label{Omega2mu0}
\end{eqnarray}
Substituting the reduced formula of $\Omega^{(2)}$ [Eq.\ (\ref{Omega2_reduced})] into this and using 
the $\mu^{(0)}$-derivatives of the Fermi--Dirac functions [Eq.\ (\ref{mu_derivative})], we obtain the reduced formula for $\mu^{(2)}$ as
\begin{widetext}
\begin{eqnarray}
\mu^{(2)} \sum_p f_p^- f_p^+  
&=& \sum_{p,q}^{\text{denom.}\neq 0} \frac{| F_{pq} |^2 f_p^- f_q^+(f_p^+ - f_q^-)}{\epsilon_p-\epsilon_q}
+ \sum_{p,q,r}^{\text{denom.}\neq 0} \frac{ (F_{qp} \langle pr||qr\rangle + \langle qr||pr\rangle F_{pq} ) f_p^- f_q^+ f_r^- f_r^+ }{\epsilon_p-\epsilon_q}
\nonumber\\&& 
+ \frac{1}{4} \sum_{p,q,r,s}^{\text{denom.}\neq 0} \frac{| \langle pq || rs \rangle |^2 f_p^- f_q^- f_r^+ f_s^+(f_p^+ + f_q^+ - f_r^- - f_s^-) }{\epsilon_p+\epsilon_q-\epsilon_r-\epsilon_s} 
- \frac{\beta}{2} \sum_{p,q}^{\text{denom.}= 0} {| F_{pq} |^2 f_p^- f_q^+ (f_p^+ - f_q^-)}
\nonumber\\&& 
- \frac{\beta}{2} \sum_{p,q,r}^{\text{denom.}= 0} {(F_{qp} \langle pr||qr\rangle + \langle qr||pr\rangle F_{pq} ) f_p^- f_q^+ f_r^- f_r^+ }
- \frac{\beta}{8} \sum_{p,q,r,s}^{\text{denom.}= 0} {| \langle pq || rs \rangle |^2 f_p^- f_q^- f_r^+ f_s^+(f_p^+ + f_q^+ - f_r^- - f_s^-)} 
\nonumber\\&& 
+ \beta \mu^{(1)} \sum_p F_{pp}f_p^- f_p^+ (f_p^+ - f_p^-)
+ \beta \mu^{(1)} \sum_{p,q} \langle pq || pq \rangle f_p^- f_p^+ f_q^-  f_q^+
- \frac{\beta}{2} \left(\mu^{(1)}\right)^2 \sum_p f_p^- f_p^+ (f_p^+ - f_p^-) . \label{mu2_reduced} 
\end{eqnarray}
\end{widetext}
Dropping $\mu^{(1)}$ as a fixed variable is permitted and leads to the same result.

\subsection{Internal energy}

Collecting the second-order terms in $U$ [Eq.\ (\ref{Udef2})], we have the sum-over-states formula
for $U^{(2)}$ that reads
\begin{eqnarray}
U^{(2)} &=& \langle E_I^{(2)} \rangle -{\beta} \langle E_I^{(1)} (E_I^{(1)}-\mu^{(1)} N_I) \rangle
+{\beta} \langle E_I^{(1)} \rangle  \langle E_I^{(1)}-\mu^{(1)} N_I \rangle 
\nonumber\\&& 
-{\beta} \langle  E_I^{(0)} (E_I^{(2)}-\mu^{(2)}N_I) \rangle
+{\beta} \langle E_I^{(0)} \rangle \langle E_I^{(2)}-\mu^{(2)}N_I\rangle 
\nonumber\\&& 
+\frac{\beta^2}{2} \langle E_I^{(0)} (E_I^{(1)}-\mu^{(1)} N_I)^2\rangle
-\frac{\beta^2}{2} \langle E_I^{(0)} \rangle \langle (E_I^{(1)}-\mu^{(1)} N_I)^2 \rangle  
\nonumber\\&& 
-{\beta^2} \langle E_I^{(0)} (E_I^{(1)}-\mu^{(1)} N_I) \rangle\langle E_I^{(1)}-\mu^{(1)} N_I \rangle
\nonumber\\&& 
+{\beta^2} \langle E_I^{(0)} \rangle \langle E_I^{(1)}-\mu^{(1)} N_I \rangle^2 . \label{U2_SoS}
\end{eqnarray}

Comparing this with the $\beta$-derivative of the sum-over-states formula of $\Omega^{(2)}$ [Eq.\ (\ref{Omega2_start})], we find
\begin{eqnarray}
\frac{\partial}{\partial \beta} \left( \beta \Omega^{(2)} \right) 
&=& -\frac{\partial}{\partial \beta} \left( \frac{\Xi^{(2)}}{\Xi^{(0)}} \right) + \frac{1}{2} \frac{\partial}{\partial \beta} \left(\frac{\Xi^{(1)}}{\Xi^{(0)}} \right)^2
\\
&=& U^{(2)} -\mu^{(2)}\bar{N} - \beta \frac{\partial \mu^{(2)} }{\partial \beta} \bar{N},
\end{eqnarray}
where Eqs.\ (\ref{1NminusN2}) and (\ref{2NminusN1}) were used in the last equality. 
Substituting the reduced formula of $\Omega^{(2)}$ [Eq.\ (\ref{Omega2_reduced})] into this and using the $\beta$-derivatives of the 
Fermi--Dirac functions [Eq.\ (\ref{beta_derivative})], we obtain the reduced formula for $U^{(2)}$ as
\begin{widetext}
\begin{eqnarray}
U^{(2)} 
&=& \frac{\partial}{\partial \beta}\left( \beta \Omega^{(2)}\right) + \mu^{(2)} \bar{N} + \beta \frac{\partial \mu^{(2)}}{\partial \beta} \bar{N}
= \Omega^{(2)} + \mu^{(2)}\bar{N} + \beta \left( \frac{\partial \Omega^{(2)}}{\partial \beta} \right)_{\mu^{(0)},\,\mu^{(1)},\,\mu^{(2)}} \label{U2fromOmega2}\\
&=& \sum_{p,q}^{\text{denom.}\neq 0} \frac{| F_{pq} |^2 f_p^- f_q^+}{\epsilon_p-\epsilon_q}
+ \frac{1}{4} \sum_{p,q,r,s}^{\text{denom.}\neq 0} \frac{| \langle pq || rs \rangle |^2 f_p^- f_q^- f_r^+ f_s^+}{\epsilon_p+\epsilon_q-\epsilon_r-\epsilon_s} 
- {\beta} \sum_{p,q}^{\text{denom.}= 0} {| F_{pq} |^2 f_p^- f_q^+}
- \frac{\beta}{4} \sum_{p,q,r,s}^{\text{denom.}= 0} {| \langle pq || rs \rangle |^2 f_p^- f_q^- f_r^+ f_s^+} 
\nonumber\\&& 
+ 2 \beta \mu^{(1)} \sum_p F_{pp}f_p^- f_p^+ - {\beta} \left(\mu^{(1)}\right)^2 \sum_p f_p^- f_p^+
- \beta \sum_{p,q}^{\text{denom.}\neq 0} \frac{| F_{pq} |^2 f_p^- f_q^+ (\epsilon_pf_p^+ - \epsilon_qf_q^-)}{\epsilon_p-\epsilon_q}
\nonumber\\&& 
- \beta \sum_{p,q,r}^{\text{denom.}\neq 0} \frac{ (F_{qp} \langle pr||qr\rangle + \langle qr||pr\rangle F_{pq} )  f_p^- f_q^+  (\epsilon_r f_r^- f_r^+) }{\epsilon_p-\epsilon_q}
- \frac{\beta}{4} \sum_{p,q,r,s}^{\text{denom.}\neq 0} \frac{| \langle pq || rs \rangle |^2 f_p^- f_q^- f_r^+ f_s^+
(\epsilon_pf_p^+ + \epsilon_qf_q^+ - \epsilon_rf_r^- - \epsilon_sf_s^-) }{\epsilon_p+\epsilon_q-\epsilon_r-\epsilon_s} 
\nonumber\\&& 
+ \frac{\beta^2}{2} \sum_{p,q}^{\text{denom.}= 0} {| F_{pq} |^2 f_p^- f_q^+ (\epsilon_p f_p^+ - \epsilon_q f_q^-)}
+ \frac{\beta^2}{2} \sum_{p,q,r}^{\text{denom.}= 0} {(F_{qp} \langle pr||qr\rangle + \langle qr||pr\rangle F_{pq} )  f_p^- f_q^+ (\epsilon_r f_r^-  f_r^+) }
\nonumber\\&& 
+ \frac{\beta^2}{8} \sum_{p,q,r,s}^{\text{denom.}= 0} {| \langle pq || rs \rangle |^2 f_p^- f_q^- f_r^+ f_s^+(\epsilon_p f_p^+ + \epsilon_q f_q^+ - \epsilon_r f_r^- - \epsilon_s f_s^-)} 
- \beta^2 \mu^{(1)} \sum_p F_{pp}f_p^- f_p^+ (\epsilon_p f_p^+ - \epsilon_p f_p^-)
\nonumber\\&& 
- \beta^2 \mu^{(1)} \sum_{p,q} \langle pq || pq \rangle   f_p^- f_p^+ (\epsilon_q f_q^-  f_q^+)
+ \frac{\beta^2}{2} \left(\mu^{(1)}\right)^2 \sum_p  f_p^-f_p^+ (\epsilon_p f_p^+ - \epsilon_pf_p^-) 
+ \beta \mu^{(2)} \sum_p \epsilon_p f_p^-f_p^+ ,
\label{U2_reduced} 
\end{eqnarray}
\end{widetext}
where $\mu^{(0)}$ and $\mu^{(1)}$ may be dropped as fixed variables (giving the same result), 
but $\mu^{(2)}$ must be held fixed. If the last two terms of $\Omega^{(2)}$ in Eq.\ (\ref{Omega2_reduced}) are consolidated into one,
the $\beta$-derivative of $\mu^{(1)}$ must be considered.
The $\mu^{(0)}$ factors in Eq.\ (\ref{beta_derivative}) accompanying every $\epsilon_p$ can be shown to 
cancel with one another and disappear by virtue of Eq.\ (\ref{2NminusN1}).

\section{Numerical verification\label{verification}} 

\begin{table*} 
\caption{\label{tab:only} Comparison of the second-order corrections to grand potential ($\Omega^{(2)}$), chemical potential ($\mu^{(2)}$), 
and internal energy ($U^{(2)}$) obtained with the $\lambda$-variation, sum-over-states formulas,
and reduced (sum-over-orbitals) formulas  
as a function of temperature ($T$) for the hydrogen fluoride molecule (0.9168~\AA) in the STO-3G basis set.}
\begin{ruledtabular}
\begin{tabular}{lddddddddd}
& \multicolumn{3}{c}{$\Omega^{(2)} / E_{\text{h}}$} &\multicolumn{3}{c}{$\mu^{(2)} / E_{\text{h}}$} & \multicolumn{3}{c}{$U^{(2)} / E_{\text{h}}$} \\ \cline{2-4} \cline{5-7} \cline{8-10}
&\multicolumn{1}{c}{$\lambda$-variation\footnotemark[1]}  
&\multicolumn{1}{c}{SoS\footnotemark[2]}  
&\multicolumn{1}{c}{Reduced} 
&\multicolumn{1}{c}{$\lambda$-variation\footnotemark[1]}  
&\multicolumn{1}{c}{SoS\footnotemark[2]}  
&\multicolumn{1}{c}{Reduced} 
&\multicolumn{1}{c}{$\lambda$-variation\footnotemark[1]}  
&\multicolumn{1}{c}{SoS\footnotemark[2]}  
&\multicolumn{1}{c}{Reduced}  \\
{$T /~\text{K}$} 
&\multicolumn{1}{c}{Eq.~(\ref{lambda})}  
&\multicolumn{1}{c}{Eq.~(\ref{Omega2_SoS})}  
&\multicolumn{1}{c}{Eq.~(\ref{Omega2_reduced})} 
&\multicolumn{1}{c}{Eq.~(\ref{lambda})}  
&\multicolumn{1}{c}{Eq.~(\ref{mu2_SoS})}  
&\multicolumn{1}{c}{Eq.~(\ref{mu2_reduced})} 
&\multicolumn{1}{c}{Eq.~(\ref{lambda})}  
&\multicolumn{1}{c}{Eq.~(\ref{U2_SoS})}  
&\multicolumn{1}{c}{Eq.~(\ref{U2_reduced})}  \\ \hline
$10^4$ & -0.4324 & \multicolumn{1}{r}{$\dots$\footnotemark[3]}& -0.4324 &  0.0415 & \multicolumn{1}{r}{$\dots$\footnotemark[3]} & 0.0415 & -0.0173 & \multicolumn{1}{r}{$\dots$\footnotemark[3]}  & -0.0173  \\
$10^5$ & -2.5815 & -2.5815 & -2.5815 & 0.2320 & 0.2320 & 0.2320 & 0.0984 & 0.0984 & 0.0984 \\
$10^6$ & -0.9643 & -0.9643 & -0.9643 &  0.0851 & 0.0851 & 0.0851 & -0.2198 & -0.2198 & -0.2198 \\
$10^7$ & -0.1970 & -0.1970 & -0.1970 &  0.0178 & 0.0177 & 0.0177 & -0.0326 & -0.0326 & -0.0326 \\
$10^8$ & -0.0276 & -0.0276 & -0.0276 &  0.0025 & 0.0025 & 0.0025 & -0.0054 & -0.0054 & -0.0054 \\
\end{tabular}
\footnotetext[1]{Taken from Jha and Hirata.\cite{JhaHirata}}
\footnotetext[2]{Obtained using the sum-over-states formulas with the first- and second-order Hirschfelder--Certain perturbation corrections to energy computed 
by the $\lambda$-variation method using forward seven- through nine-point finite differences.}
\footnotetext[3]{Numerical instability due to roundoff errors.}
\end{ruledtabular}
\end{table*}

Table \ref{tab:only} lists the numerical values of $\Omega^{(2)}$, $\mu^{(2)}$, and $U^{(2)}$ of the hydrogen fluoride molecule (0.9168~\AA) in the STO-3G basis set
as a function of temperature.\cite{Kou} They were computed by three different methods:\ the ``$\lambda$-variation'' method,\cite{JhaHirata} which computes $X^{(n)}$ ($X = \Omega$,
$\mu$, or $U$) by a finite-difference approximation to the $\lambda$-derivatives [Eq.\ (\ref{lambda})], the sum-over-states (``SoS'') analytical formulas, and the sum-over-orbitals (``reduced'') analytical 
formulas. They agree numerically exactly with one another in a wide range of temperature, attesting to their correctness (for a nondegenerate ground state).

See our previous paper\cite{HirataJha} for the numerical verification of $\Omega^{(1)}$, $\mu^{(1)}$, and $U^{(1)}$.

\section{Conclusions}

Unlike thermodynamics of vibrations, rotations, and translations, which are well understood,
finite-temperature perturbation theory for electrons has been plagued with controversy and confusion.\cite{kohn,luttingerward,hirata_KL,SANTRA,WhiteChan} The confusion (if not the controversy itself) 
is traced at least partly to the fact that electrons are charged particles.\cite{JhaHirata} The chemical potential $\mu$ must be adjusted at each perturbation order to ensure the
overall electroneutrality of the macroscopic system being described. 

There are more than one way to make such adjustments, but expanding $\mu$ as well as $\Omega$, $U$, and $S$ in a consistent perturbation series 
may be the most satisfactory route because it conforms to the canonical definition of a perturbation theory [Eq.\ (\ref{lambda})]. 
Only such canonical perturbation theories (MPPT for nondegenerate states,\cite{shavitt} HCPT for degenerate states,\cite{Hirschfelder} and the Feynman--Dyson
perturbation series for many-body Green's function theory\cite{Hirata2017}) tend to survive the test of time. 
The finite-temperature perturbation theory introduced here expands $\Omega$, $\mu$, $U$ and $S$ on an equal footing
in accordance with Eq.\ (\ref{lambda}).

In a previous article,\cite{HirataJha} we presented the sum-over-states and reduced (sum-over-orbitals) analytical formulas of these properties 
at the first order of such a finite-temperature many-body perturbation theory.\cite{JhaHirata} 
In this article, we extended this theory to the second order.
The reduced analytical formula for $\Omega^{(2)}$ differs from the one found in the 
textbooks.\cite{thouless1972quantum,mattuck1992guide,march1995many,Fetter,SANTRA} 
To the authors' knowledge, the analytical formulas for $\mu^{(2)}$ and $U^{(2)}$ have been unknown  
until they are presented in this article for the first time. 
They reproduce the benchmark data\cite{JhaHirata} numerically exactly.

Equally important to these analytical formulas is the general and transparent (if tedious) strategy of their derivation.
It is time-independent, nondiagrammatic, and algebraic, using elementary combinatorics and calculus with the only nontrivial
step being the sum rules of HCPT. 
The latter is inevitable because the perturbation corrections to energies that are being thermally averaged come from HCPT,
which is apparently hidden from view in the quantum-field-theoretical\cite{thouless1972quantum,mattuck1992guide,march1995many,Fetter} or density-matrix-based\cite{SANTRA} derivations
of the finite-temperature perturbation theory in many textbooks that differs from this work.
We hope that our derivation will serve to justify and perhaps generalize the diagrammatic logic as well as to 
sharpen and answer\cite{Hirata_KLredux} the question that is the basis of the aforementioned controversy.\cite{kohn,luttingerward}
We  expect that this work will also help clarify the precise relationship between the finite-temperature many-body perturbation theory\cite{HirataJha} and
other finite-temperature theories.\cite{sanyal,Mandal2,mandal,Zgid,Zgid2,WhiteChan2,Harsha1,Harsha2,Nooijen}

\section{Data Availability Statement}
The data that supports the findings of this study are available within the article.

\acknowledgments

This work was supported by the Center for Scalable, Predictive methods for Excitation and Correlated phenomena (SPEC), which is funded by 
the U.S. Department of Energy, Office of Science, Office of Basic Energy Sciences, 
Chemical Sciences, Geosciences, and Biosciences Division, as a part of the Computational Chemical Sciences Program
and also by the U.S. Department of Energy, Office of Science, Office of Basic Energy Sciences under Grant No.\ DE-SC0006028.
We sincerely thank Dr.\ Garnet K.-L. Chan, Dr.\ Peter J. Knowles, Mr.\ Jonathon Misiewicz, Dr.\ Debashis Mukherjee, Dr.\ Marcel Nooijen, Dr.\ Mark R. Pederson, 
Dr.\ Robin Santra, Dr.\ Samuel B. Trickey, and Dr.\ Alec F. White for helpful discussions.

\appendix

\section{Boltzmann-sum identities\label{BoltzmannIdentities}}

Here, we list all the algebraic identities needed to derive 
the reduced analytical formulas in this article. 
These identities can be inferred from elementary combinatorics, and are not limited to the thermodynamics application insofar
as an exponentially weighted sum is taken over all $2^n$ occupancies of $n$ slots. Equation (\ref{identity0}) may be considered as Identity 0.

Index $I$ runs over all $2^n$ Slater determinants, where $n$ is the number of spinorbitals.
Index $i$ in $\sum_i^I$ goes through all spinorbitals occupied by an electron in the $I$th determinant.
Index $a$ in $\sum_a^I$ runs over all spinorbitals unoccupied in the $I$th determinant. 
We use letters $i$, $j$, and $k$ for an occupied spinorbital in the $I$th determinant, $a$ and $b$ for an unoccupied spinorbital in the same, and 
$p$, $q$, $r$, and $s$ for a general spinorbital.

The first identity is
\begin{widetext}
\begin{eqnarray}
\frac{ \sum_{I_1} \sum_i^{I_1} a_i e^{\nu_i} + \sum_{I_2}\sum_{i<j}^{I_2} (a_i + a_j) e^{\nu_i} e^{\nu_j}+ \sum_{I_3} \sum_{i<j<k}^{I_3} (a_i + a_j + a_k) e^{\nu_i} e^{\nu_j}e^{\nu_k}  + \dots }{\prod_p (1+ e^{\nu_p}) } = \sum_p a_p f_p^-  \,\,;\,\,(\text{Identity I}),\label{identity1}
\end{eqnarray}
where $\sum_{I_1}$ sums over all one-electron ($I_1$) Slater determinants, $\sum_{I_2}$ over all two-electron ($I_2$) determinants, etc.\ with ``$\dots$'' includes
the sum over up to all $n$-electron determinants (though there is only one). $\sum_{i}^{I_1}$ means that $i$ runs over all spinorbitals occupied in $I_1$, while
$p$ in $\prod_p$ goes through all spinorbitals. $a_p$ is a complex number and $f_p^- = (1+e^{-\nu_p})^{-1}$.
The second identity reads
\begin{eqnarray}
&& \frac{ \sum_{I_1}  \sum_i^{I_1} a_i a^\prime_i e^{\nu_i} + \sum_{I_2} \sum_{i<j}^{I_2} (a_i + a_j)(a^\prime_i + a^\prime_j) e^{\nu_i}e^{\nu_j}  + \sum_{I_3}\sum_{i<j<k}^{I_3} (a_i+a_j+a_k)(a^\prime_i+a^\prime_j+a^\prime_k)e^{\nu_i}e^{\nu_j}e^{\nu_k} 
+ \dots  }{\prod_p (1+ e^{\nu_p})} 
 \nonumber\\&& = \sum_p a_pa^\prime_p f_p^-f_p^+ +\left\{  \sum_{p} a_p f_p^- \right\}\left\{ \sum_p a^\prime_p f_p^-\right\} \,\,;\,\,(\text{Identity II}),\label{identity2}
\end{eqnarray}
where $a_p$ and $a^\prime_p$ are two independent complex numbers, and $f_p^+ = 1-f_p^-$. The third identity is
\begin{eqnarray}
\frac{ \sum_{I_2} \sum_{i<j}^{I_2} b_{ij} e^{\nu_i}e^{\nu_j}  + \sum_{I_3} \sum_{i<j<k}^{I_3} (b_{ij} + b_{ik} + b_{jk} )e^{\nu_i}e^{\nu_j}e^{\nu_k} + \dots  }{\prod_p (1+ e^{\nu_p})} = \frac{1}{2} \sum_{p,q} b_{pq} f_p^- f_q^- \,\,;\,\,(\text{Identity III}),\label{identity3}
\end{eqnarray}
where $b_{pq}$ is real and $b_{pq} = b_{qp}$ as well as $b_{pp} = 0$. The fourth identity is
\begin{eqnarray}
&& \frac{ \sum_{I_2}  \sum_{i<j}^{I_2} (a_i + a_j) b_{ij} e^{\nu_i}e^{\nu_j}  +\sum_{I_3} \sum_{i<j<k}^{I_3} (a_i + a_j+a_k) (b_{ij} + b_{ik} + b_{jk} )e^{\nu_i}e^{\nu_j}e^{\nu_k} + \dots }{\prod_p (1+ e^{\nu_p})} 
 \nonumber\\&& = \frac{1}{2} \sum_{p,q} (a_p f_p^+ + a_q f_q^+) b_{pq} f_p^- f_q^- + \frac{1}{2} \left\{ \sum_p a_p f_p^- \right\} \left\{ \sum_{p,q} b_{pq} f_p^- f_q^-\right\}  \,\,;\,\,(\text{Identity IV}), \label{identity4}
\end{eqnarray}
where $a_p$ is complex, $b_{pq}$ is real, $b_{pq} = b_{qp}$, and $b_{pp}=0$. 

Identities I through IV were already presented in our earlier paper.\cite{HirataJha}

A new identity involves spinorbitals that are occupied (labeled by $i$ and $j$) and unoccupied (labeled by $a$), and it reads
\begin{eqnarray}
 \frac{\sum_{I_{1}} \sum_{i,a}^{I_{1}} b_{ia}  e^{\nu_i} + \sum_{I_{2}} \sum_{i<j,a}^{I_{2}} ( b_{ia} + b_{ja} ) e^{\nu_i}e^{\nu_j} +
  \dots }{\prod_p (1+ e^{\nu_p})} 
 = \sum_{p}\sum_{q \neq p} b_{pq} f_p^- f_q^+  \,\,;\,\,(\text{Identity V}),\label{identity5}
\end{eqnarray}
where $b_{pq}$ is complex. The restriction $q \neq p$ in the right-hand side comes from the fact that no spinorbital can be simultaneously occupied and unoccupied in any Slater determinant.

The sixth identity is
\begin{eqnarray}
&& \frac{\sum_{I_2} \sum_{i<j}^{I_2} ( b_{ij} )^2 e^{\nu_i} e^{\nu_j} + \sum_{I_3} \sum_{i<j<k}^{I_3} ( b_{ij} + b_{ik} + b_{jk} )^2 e^{\nu_i} e^{\nu_j} e^{\nu_k} + \dots }{\prod_p (1+ e^{\nu_p})} \nonumber\\
&&= \frac{1}{2} \sum_{p,q} (b_{pq})^2 f_p^- f_q^-
+\sum_{p,q}\sum_{r \neq q} b_{pq}b_{pr}f_p^- f_q^- f_r^- + \frac{1}{4} \sum_{p,q,r,s}^\text{no coinc.} b_{pq}b_{rs} f_p^- f_q^- f_r^- f_s^- \,\,;\,\,(\text{Identity VI}),\label{identity6}
\end{eqnarray}
where $b_{pq}$ is real and $b_{pq} = b_{qp}$ as well as $b_{pp} = 0$. Superscript ``no coinc.''\ stands for excluding all cases where 
two or more indices are coincident, namely, $p=r$, $p=s$, $q=r$, or $q=s$ ($p=q$ or $r=s$ is also excluded, but this does not have
to be explicitly stated because the corresponding summands are zero, i.e., $b_{pp} = b_{rr} = 0$). 

The seventh identity reads
\begin{eqnarray}
\frac{\sum_{I_2} \sum_{i,j,a}^{I_2} c_{ija}  e^{\nu_i}e^{\nu_j} + \sum_{I_3} \sum_{i,j<k,a}^{I_3} ( c_{ija} + c_{ika} ) e^{\nu_i}e^{\nu_j}e^{\nu_k} +  \dots }{\prod_p (1+ e^{\nu_p})} = \sum_{p ,q}\sum_{r \neq p} c_{pqr} f_p^- f_q^-  f_r^+  \,\,;\,\,(\text{Identity VII}),\label{identity7}
\end{eqnarray}
where $c_{pqr}$ is complex and $c_{ppr} = c_{pqq} = 0$. The eighth is
\begin{eqnarray}
&& \frac{\sum_{I_2} \sum_{i,j,a}^{I_2} | c_{ija} |^2 e^{\nu_i}e^{\nu_j} 
+ \sum_{I_3} \sum_{i,j<k,a}^{I_3}  | c_{ija} + c_{ika} |^2 e^{\nu_i}e^{\nu_j}e^{\nu_k} 
+  \dots }{\prod_p (1+ e^{\nu_p})} \nonumber\\
&&= \sum_{p}\sum_{r \neq p}  \left| \sum_{q} c_{pqr} f_q^- \right|^2 f_p^- f_r^+ + \sum_{p,q}\sum_{r \neq p}|c_{pqr}|^2 f_p^- f_r^+   f_q^-  f_q^+ \,\,;\,\,(\text{Identity VIII}),\label{identity8}
\end{eqnarray}
where $c_{pqr}$ is complex and $c_{ppr} = c_{pqq} = 0$.  

The ninth identity for real $d_{ijab}$ is
\begin{eqnarray}
\frac{\sum_{I_2}  \sum_{i < j,a<b}^{I_2} d_{ijab} e^{\nu_i}e^{\nu_j} 
+ \sum_{I_3} \sum_{i < j<k,a<b}^{I_3} (d_{ijab} + d_{ikab}+ d_{jkab}) e^{\nu_i}e^{\nu_j}e^{\nu_k} 
+ \dots  }{\prod_p (1+ e^{\nu_p})}  
 = \frac{1}{4} \sum_{p,q,r,s}^\text{no coinc.} d_{pqrs} f_p^- f_q^- f_r^+ f_s^+ \,\,;\,\,(\text{Identity IX}),\label{identity9}
\end{eqnarray}
where $d_{pqrs} = d_{qprs} = d_{pqsr} = d_{qpsr}$ and $d_{pprs} = d_{pqrr} = 0$ are presumed. 
Superscript ``no coinc.''\ stands for excluding all cases where two or more indices are coincident, namely, $p=r$, $p=s$, $q=r$,  or $q=s$, 
which ultimately arises from the fact that no spinorbital is simultaneously occupied and unoccupied in any Slater determinant
(the exclusion of $p=q$ or $r=s$ is effected by $d_{pprs} = d_{pqrr} = 0$). Remarkably, such restrictions are systematically lifted 
in the final reduced formula of $\Omega^{(2)}$ (see Appendix \ref{DerivationOmega2}).

\end{widetext}

\section{Hirschfelder--Certain sum rules\label{Sumrules}}

For a state whose zeroth-order energy is degenerate,
a perturbation correction to energy $E_I^{(n)}$ cannot be written in a closed expression of
molecular integrals (or diagrammatically); it is defined only 
procedurally by HCPT.\cite{Hirschfelder} This procedure, in turn, involves diagonalization of a matrix in the degenerate 
subspace, whose outcome is generally not expressible as a sum-of-products of integrals. 

Nevertheless, the sum of $E_I^{(n)}$ in a degenerate subspace can be written 
in a closed formula of integrals, as shown below. Since, in the zeroth-order thermal average, 
they are summed with an equal weight, $e^{-\beta E_I^{(0)} + \beta \mu^{(0)} N_I}$, dictated by $E_I^{(0)}$ and $N_I$ (which are
common within the degenerate subspace), 
it is the sums of $E_I^{(n)}$ in degenerate subspaces (rather than individual $E_I^{(n)}$) that we need in order to 
correctly evaluate the average. Here, we derive and document such sum rules.  

For a nondegenerate state, HCPT reduces to MPPT, giving the closed formula for $E_I^{(1)}$,\cite{szabo,HirataJha}
\begin{eqnarray}
E_I^{(1)} &=& \sum_{i}^I H_{ii}^{\text{core}} + \sum_{i < j}^I \langle ij || ij \rangle - \sum_i^I \epsilon_i, \label{E1nondeg}
\end{eqnarray}
which is equal to $\langle \Phi_I | \hat{V}|\Phi_I \rangle$ evaluated by the Slater--Condon rules, where $\Phi_I$ is the $I$th Slater determinant and $\hat{V}$ is the perturbation
operator [Eqs.\ (\ref{partitioning0}) and (\ref{partitioning})].

For a degenerate subspace, we have\cite{szabo,HirataJha}
\begin{eqnarray}
\sum_I^{\text{degen.}} E_I^{(1)} &=& \sum_I^{\text{degen.}} \left\{ \sum_{i}^I H_{ii}^{\text{core}} + \sum_{i < j}^I \langle ij || ij \rangle - \sum_i^I \epsilon_i \right\} , \label{E1sumrule}
\end{eqnarray}
where ``degen.''\ means that $I$ runs over all Slater determinants in the degenerate subspace, sharing the same $E_I^{(0)}$ and $N_I$. 

Equation (\ref{E1sumrule}) can be rationalized as follows. According to Eq.\ (37) of Hirschfelder and Certain,\cite{Hirschfelder} $E_I^{(1)}$ within a degenerate subspace are the eigenvalues of the matrix $\bm{V}$ whose element is $V_{IJ} = \langle \Phi_I |\hat{V} | \Phi_J \rangle$ (where $\Phi_I$ and $\Phi_J$ are two Slater determinants in the degenerate subspace). 
Owing to the similarity invariance of trace, 
the sum of the eigenvalues (the left-hand side of the above equation) is equal to the sum of the diagonal elements (the right-hand side), each of which is readily evaluated by the Slater--Condon rules.\cite{szabo,HirataJha} 

For a nondegenerate state, $E_I^{(2)}$ has a well-known formula,\cite{szabo,shavitt}
\begin{eqnarray}
E_I^{(2)} &=&  \sum_{i,a}^I \frac{ \left |H^{\text{core}}_{ia} + \sum_j^I \langle ij||aj \rangle \right |^2 }{\epsilon_i - \epsilon_a} 
+ \sum_{i < j,a < b}^I \frac{|\langle ij||ab\rangle|^2}{\epsilon_i + \epsilon_j - \epsilon_a - \epsilon_b}, \nonumber\\  \label{E2nondeg}  
\end{eqnarray}
where $i$ and $j$ run over spinorbitals occupied in the $I$th Slater determinant, while $a$ and $b$ refer to spinorbitals unoccupied in the same determinant.
The first term is identified as the non-HF term\cite{shavitt} with the numerator factor recognized as the $ia$-th element of the zero-temperature Fock matrix\cite{szabo} of the $I$th state. 

The sum rule for $E_I^{(2)}$ within a degenerate subspace is
\begin{widetext}
\begin{eqnarray}
\sum_I^{\text{degen.}} E_I^{(2)} &=& \sum_I^{\text{degen.}}\left\{  \sum_{i,a}^{I,\,\text{denom.}\neq 0} \frac{ \left |H^{\text{core}}_{ia} + \sum_j^I \langle ij||aj \rangle \right |^2 }{\epsilon_i - \epsilon_a} 
+ \sum_{i < j,a < b}^{I,\,\text{denom.}\neq 0} \frac{|\langle ij||ab\rangle|^2}{\epsilon_i + \epsilon_j - \epsilon_a - \epsilon_b} \right\} , \label{E2sumrule}
\end{eqnarray}
\end{widetext}
where ``$I,\,\text{denom.}\neq 0$'' means $i,j$ ($a,b$) run over spinorbitals occupied (unoccupied) in the $I$th state excluding the case with a vanishing denominator. 
In other words, $E_I^{(2)}$ for a degenerate state 
accumulates the usual MPPT-type second-order corrections [Eq.\ (\ref{E2nondeg})] only from outside the degenerate subspace. 
This sum rule can be justified by Eqs.\ (49), (56), and (57) of Hirschfelder and Certain\cite{Hirschfelder} as well as the trace invariance.

The sum rule for $(E_I^{(1)})^2$ within a degenerate subspace reads
\begin{eqnarray}
\sum_I^{\text{degen.}} \left( E_I^{(1)}\right)^2 &=& \sum_I^{\text{degen.}} \left( E_I^{\text{0e}}  + E_I^{\text{1e}} + E_I^{\text{2e}} \right), \label{E1E1sumrule}
\end{eqnarray}
with
\begin{eqnarray}
E_I^{\text{0e}} &=& \left\{ \sum_{i}^I H_{ii}^{\text{core}} + \sum_{i < j}^I \langle ij || ij \rangle - \sum_i^I \epsilon_i \right\}^2, \label{E0_notyet}\\
E_I^{\text{1e}} &=& \sum_{i,a}^{I,\,\text{denom.}= 0} { \left |H^{\text{core}}_{ia} + \sum_j^I \langle ij||aj \rangle \right |^2 }, \label{E1_notyet}\\
E_I^{\text{2e}} &=& \sum_{i < j,a < b}^{I,\,\text{denom.}= 0} {|\langle ij||ab\rangle|^2}, \label{E2_notyet}
\end{eqnarray}
where ``$I, \text{denom.}=0$''  in $E_I^{\text{1e}}$  means that $i$ and $a$ run over all occupied and unoccupied spinorbitals in the $I$th state 
that satisfy $\epsilon_i - \epsilon_a = 0$. The same in $E_I^{\text{2e}}$ demands $\epsilon_i + \epsilon_j - \epsilon_a - \epsilon_b = 0$. 
Therefore, in contrast to the sum rule for $E_I^{(2)}$ [Eq.\ (\ref{E2sumrule})], $E_I^{\text{1e}}$ and $E_I^{\text{2e}}$ 
accumulate denominatorless MPPT-type corrections only from inside the degenerate subspace. 
The unlinked-diagram term, $E_I^{\text{0e}}$, does not have such restrictions. 

This sum rule is rationalized as follows. 
Since $E_I^{(1)}$ in the degenerate subspace are the eigenvalues of the matrix $\bm{V}$,
$(E_I^{(1)})^2$ are the eigenvalues of the matrix $\bm{V}^2$. The trace invariance of the latter implies
that the sum of $(E_I^{(1)})^2$ is equal to the sum of the diagonal elements of $\bm{V}^2$, which can then be expanded as
\begin{eqnarray}
 \sum_I^{\text{degen.}} \left( E_I^{(1)}\right)^2 
 &=& \sum_I^{\text{degen.}} \langle \Phi_I |\hat{V} |\Phi_I \rangle \langle\Phi_I |\hat{V} | \Phi_I \rangle \nonumber \\
&&+ \sum_I^{\text{degen.}} \sum_{i,a}^{I,\,\text{denom.}= 0} \langle \Phi_I |\hat{V} |\Phi^a_i \rangle \langle\Phi^a_i |\hat{V} | \Phi_I \rangle \nonumber \\
&& + \sum_I^{\text{degen.}} \sum_{i<j,a<b}^{I,\,\text{denom.}= 0} \langle \Phi_I |\hat{V} |\Phi^{ab}_{ij} \rangle \langle\Phi^{ab}_{ij} |\hat{V} | \Phi_I \rangle, 
\nonumber\\
\end{eqnarray} 
where $\Phi^a_i$ runs over all degenerate Slater determinants that are one-electron replacement from $\Phi_I$ 
(an electron in the $i$th spinorbital occupied in $\Phi_I$ 
is promoted to the $a$th spinorbital unoccupied in $\Phi_I$ with $\epsilon_a = \epsilon_i$). Similarly, 
$\Phi^{ab}_{ij}$ runs over all degenerate Slater determinants that are two-electron replacement from $\Phi_I$
(with $\epsilon_a + \epsilon_b = \epsilon_i + \epsilon_j$). 
The first term corresponds to an unlinked-diagram contribution and is identified as $E_I^{\text{0e}}$ [Eq.\ (\ref{E0_notyet})], whereas
the second and third terms are linked and are evaluated by the Slater--Condon rules\cite{szabo} as $E_I^{\text{1e}}$ [Eq.\ (\ref{E1_notyet})] and $E_I^{\text{2e}}$ [Eq.\ (\ref{E2_notyet})], respectively. 

\section{Derivation of Eq.\ (\ref{Omega2_reduced})\label{DerivationOmega2}}

Using Eqs.\ (\ref{E1N}) and (\ref{NN}), we can immediately partially reduce the sum-over-states formula of $\Omega^{(2)}$ [Eq.\ (\ref{Omega2_SoS})] to
\begin{eqnarray}
\Omega^{(2)} &=& \langle E_I^{(2)} \rangle  -\frac{\beta}{2} \langle (E_I^{(1)})^2 \rangle 
+\frac{\beta}{2}  \langle E_I^{(1)} \rangle^2
- \mu^{(2)} \bar{N} 
\nonumber\\&& 
+ \beta \mu^{(1)} \sum_p F_{pp}f_p^- f_p^+ - \frac{\beta}{2} \left(\mu^{(1)}\right)^2 \sum_p f_p^- f_p^+ , \label{Omega_partial}
\end{eqnarray}
where $\langle E_I^{(1)} \rangle$ was also already simplified to Eq.\ (\ref{E1}). The remaining task is, therefore, 
to evaluate $\langle E_I^{(2)} \rangle$ and $\langle (E_I^{(1)})^2 \rangle$. 

For the purpose of reducing $\langle E_I^{(2)} \rangle$, we can pretend
that $E_I^{(2)}$ has the following closed expression for each (degenerate or nondegenerate) state:
\begin{eqnarray}
E_I^{(2)} &=& \sum_{i,a}^{I,\,\text{denom.} \neq 0} b_{ia} + \sum_{i,j,a}^{I,\,\text{denom.} \neq 0} c_{ija} - \sum_{i,a}^{I,\,\text{denom.} \neq 0} \left| \sum_j^I c_{ija}^\prime \right|^2 
\nonumber\\&&+ \sum_{i<j,a<b}^{I,\,\text{denom.} \neq 0} d_{ijab}, \label{c2}
\end{eqnarray}
with
\begin{eqnarray}
b_{ia} &=& \frac{\left| H_{ia}^{\text{core}}\right|^2} {\epsilon_i -\epsilon_a}, \\
c_{ija} &=& \frac{ H_{ai}^{\text{core}} \langle ij||aj\rangle  + \langle aj||ij\rangle  H_{ia}^{\text{core}} } {\epsilon_i -\epsilon_a}, \\
c^\prime_{ija} &=& \frac{\langle ij||aj\rangle} {(\epsilon_a -\epsilon_i)^{1/2}}, \\
d_{ijab} &=& \frac{|\langle ij||ab\rangle|^2}{\epsilon_i + \epsilon_j - \epsilon_a - \epsilon_b},
\end{eqnarray}
which encompass both Eqs.\ (\ref{E2nondeg}) and (\ref{E2sumrule}).
Since the $I$th state in Eq.\ (\ref{E2nondeg}) is nondegenerate, a vanishing denominator cannot occur and $\sum^I$ in it can, therefore,
be replaced by $\sum^{I,\,\text{denom.}\neq 0}$, justifying the ``$\text{denom.}\neq 0$'' restriction on each summation in Eq.\ (\ref{c2}).

Substituting these into Boltzmann-sum identities V, VII, VIII, and IX of Appendix \ref{BoltzmannIdentities}, we obtain
\begin{eqnarray}
\langle E_I^{(2)} \rangle &=& \sum_{p, q\neq p}^{\text{denom.}\neq 0} \frac{\left|F_{pq}\right|^2 f_p^- f_q^+}{\epsilon_p - \epsilon_q} \nonumber\\
&&+\sum_{p, q \neq p}^{\text{denom.}\neq 0}\sum_r \frac{\left|\langle pr || qr \rangle \right|^2f_p^- f_q^+ f_r^- f_r^+}{\epsilon_p - \epsilon_q} \nonumber\\
&&+ \frac{1}{4}\sum_{p,q,r,s}^{\substack{\text{no coinc.} \\ \text{denom.}\neq 0}} \frac{\left|\langle pq || rs \rangle\right|^2f_p^- f_q^- f_r^+ f_s^+}{\epsilon_p + \epsilon_q - \epsilon_r - \epsilon_s} \label{E2_reduced_1} \\
&=& \sum_{p, q}^{\text{denom.}\neq 0} \frac{\left|F_{pq}\right|^2 f_p^- f_q^+}{\epsilon_p - \epsilon_q} \nonumber\\
&&+ \frac{1}{4}\sum_{p,q,r,s}^{\text{denom.}\neq 0} \frac{\left|\langle pq || rs \rangle\right|^2f_p^- f_q^- f_r^+ f_s^+}{\epsilon_p + \epsilon_q - \epsilon_r - \epsilon_s}.\label{E2_reduced}
\end{eqnarray}
In the first term of Eq.\ (\ref{E2_reduced_1}), the restriction $q \neq p$ is effected by the nonzero denominator condition (``$\text{denom.}\neq 0$'') and can thus be lifted. 
The second term of Eq.\ (\ref{E2_reduced_1}) is absorbed by the third term to eliminate the restriction (``no coinc.'')
in Eq.\ (\ref{E2_reduced}). 
The exclusion of the contributions involving $\langle pq || pq \rangle$ or $\langle pq || qp \rangle$ is still in effect
in principle, but it is encompassed by 
the nonzero denominator condition, and is hence not explicitly demanded.
In this step, $\bm{F}$ (the finite-temperature Fock matrix minus the diagonal part of the zero-temperature Fock matrix) defined by Eq.\ (\ref{Fock}) naturally emerges.

Likewise, for the purpose of simplifying $\langle (E_I^{(1)})^2 \rangle$, we may write $(E_I^{(1)})^2$ for each (degenerate or nondegenerate) state as
\begin{eqnarray}
\left( E_I^{(1)} \right)^2 &=& E_I^{\text{0e}} + E_I^{\text{1e}} + E_I^{\text{2e}}, 
\end{eqnarray}
where
\begin{eqnarray}
E_I^{\text{0e}} &=& \left\{ \sum_{i}^I a_{i} \right\}^2  + 2 \sum_{i}^I a_i \sum_{j<k}^I b_{jk}+  \left\{ \sum_{i<j}^I b_{ij} \right\} ^2, \label{E0e_breakdown} \\
E_I^{\text{1e}} &=& \sum_{i,a}^{I,\,\text{denom.}= 0} b_{ia} + \sum_{i,j,a}^{I,\,\text{denom.}= 0}c_{ija} 
+ \sum_{i,a}^{I,\,\text{denom.}= 0} \left| \sum_j^I c_{ija}^\prime \right|^2, \nonumber\\ \label{E1e_breakdown}  \\
E_I^{\text{2e}} &=& \sum_{i<j,a<b}^{I,\,\text{denom.}= 0} d_{ijab},  \label{E2e_breakdown} 
\end{eqnarray}
and
\begin{eqnarray}
a_{i} &=& H_{ii}^{\text{core}} - \epsilon_i,  \\
b_{ij} &=& \langle ij || ij \rangle, \\
b_{ia} &=& {\left| H_{ia}^{\text{core}}\right|^2} , \\
c_{ija} &=& { H_{ai}^{\text{core}} \langle ij||aj\rangle  + \langle aj||ij\rangle  H_{ia}^{\text{core}} }, \\
c^\prime_{ija} &=& {\langle ij||aj\rangle} , \\
d_{ijab} &=& {|\langle ij||ab\rangle|^2}.
\end{eqnarray}
For a nondegenerate state, $(E_I^{(1)})^2 = (E_I^{\text{0e}})^2$. 

Following the identical strategy to obtain $\langle E_I^{(2)} \rangle$ [Eq.\ (\ref{E2_reduced})], i.e., applying 
Boltzmann-sum identities V, VII, VIII, and IX to Eqs.\ (\ref{E1e_breakdown}) and (\ref{E2e_breakdown}), we obtain
\begin{eqnarray}
\langle E_I^{\text{1e}} \rangle &=& \sum_{p, q\neq p}^{\text{denom.}= 0} {\left|F_{pq}\right|^2 f_p^- f_q^+} \nonumber\\
&&+\sum_{p, q \neq p}^{\text{denom.}= 0}\sum_r {\left|\langle pr || qr \rangle \right|^2f_p^- f_q^+ f_r^- f_r^+}, \label{E_linked_reduced_1}  \\
\langle E_I^{\text{2e}} \rangle &=& \frac{1}{4}\sum_{p,q,r,s}^{\substack{\text{no coinc.} \\ \text{denom.}= 0}} {\left|\langle pq || rs \rangle\right|^2f_p^- f_q^- f_r^+ f_s^+} , 
\end{eqnarray}
which are nothing but Eq.\ (\ref{E2_reduced_1}) stripped of the denominators.
However, adding them together, we get
\begin{eqnarray}
\langle E_I^{\text{1e}} \rangle + \langle E_I^{\text{2e}} \rangle&=& \sum_{p, q \neq p}^{\text{denom.}= 0} {\left|F_{pq}\right|^2 f_p^- f_q^+} \nonumber\\
&&+ \frac{1}{4}\sum_{p,q,r,s}^{\substack{\text{no }J\text{ or }K \\ \text{denom.}= 0}} {\left|\langle pq || rs \rangle\right|^2f_p^- f_q^- f_r^+ f_s^+},
\label{E_linked_reduced}
\end{eqnarray}
which is not Eq.\ (\ref{E2_reduced}) stripped of the denominators.
Unlike in $\langle E_I^{(2)} \rangle$, the $q \neq p$ restriction in the first term is not lifted because $\epsilon_p-\epsilon_q = 0$ can still be true even when $q \neq p$.
Similarly, the second term of $\langle E_I^{\text{1e}} \rangle$ is absorbed by $\langle E_I^{\text{2e}} \rangle$, turning the restriction on the index coincidence (``no conic.'')\ 
to only a weaker restriction (``no $J$ or $K$'') excluding the so-called $J$- and $K$-type integrals, $\langle pq || pq \rangle$ and $\langle pq || qp \rangle$, from the sum. 

Next, applying Boltzmann-sum identities II, IV, and VI to Eq.\ (\ref{E0e_breakdown}), we find
\begin{widetext}
\begin{eqnarray}
\langle E_I^{\text{0e}} \rangle &=& 
\sum_p \left( H_{pp}^{\text{core}} - \epsilon_p \right)^2 f_p^- f_p^+ 
+ \left\{ \sum_p \left( H_{pp}^{\text{core}} - \epsilon_p \right) f_p^- \right\}^2 
\nonumber\\&&
+ 2 \sum_{p,q}   \left( H_{pp}^{\text{core}} - \epsilon_p \right) f_p^+  \langle pq||pq \rangle f_p^- f_q^- 
+ \left\{ \sum_p \left( H_{pp}^{\text{core}} - \epsilon_p \right) f_p^-\right\}  \left\{ \sum_{p,q} \langle pq||pq \rangle f_p^- f_q^- \right\} 
\nonumber\\&& 
+\frac{1}{2} \sum_{p,q} |\langle pq || pq \rangle|^2 f_p^- f_q^-  
+\sum_{p,q}\sum_{r\neq q} \langle pq || pq \rangle \langle pr || pr \rangle f_p^- f_q^- f_r^- 
+ \frac{1}{4}\sum_{p,q,r,s}^{\text{no coinc.}} {\langle pq || pq \rangle \langle rs || rs \rangle f_p^- f_q^- f_r^- f_s^-} \\
&=& 
\sum_p \left( H_{pp}^{\text{core}} - \epsilon_p \right)^2 f_p^- f_p^+ 
+ \left\{ \sum_p \left( H_{pp}^{\text{core}} - \epsilon_p \right) f_p^- \right\}^2 
\nonumber\\&&
+ 2 \sum_{p,q}   \left( H_{pp}^{\text{core}} - \epsilon_p \right) f_p^+  \langle pq||pq \rangle f_p^- f_q^- 
+ \left\{ \sum_p \left( H_{pp}^{\text{core}} - \epsilon_p \right) f_p^-\right\}  \left\{ \sum_{p,q} \langle pq||pq \rangle f_p^- f_q^- \right\} 
\nonumber\\&& 
+\sum_{p}\left\{ \sum_{q} \langle pq || pq \rangle f_q^-\right\}^2 f_p^- f_p^+ 
+\frac{1}{2} \sum_{p,q} |\langle pq || pq \rangle|^2 f_p^- f_q^- f_p^+ f_q^+  
+ \frac{1}{4}\sum_{p,q,r,s} {\langle pq || pq \rangle \langle rs || rs \rangle f_p^- f_q^- f_r^- f_s^-} \\
&=& \langle E_I^{(1)}\rangle^2 + \sum_p \left|F_{pp}\right|^2 f_p^-f_p^+ 
+ \frac{1}{2} \sum_{p,q} |\langle pq || pq \rangle |^2 f_p^- f_q^- f_p^+ f_q^+. \label{E_unlinked_reduced}
\end{eqnarray}
\end{widetext}
Remarkably, the unlinked contribution $\langle E_I^{\text{0e}} \rangle$ is not canceled exactly by $\langle E_I^{(1)}\rangle^2$.
Small remainders exist, which are identified as anomalous diagrams of Kohn and Luttinger \cite{kohn} and lift the restrictions on summation indices
(both $q \neq p$ and ``no $J$ or $K$'')
in Eq.\ (\ref{E_linked_reduced}) for $\langle E_I^{\text{1e}} \rangle + \langle E_I^{\text{2e}} \rangle$.

Combining Eqs.\ (\ref{E_linked_reduced}) and (\ref{E_unlinked_reduced}), therefore, we obtain
\begin{eqnarray}
\langle (E_I^{(1)})^2\rangle &=&\langle E_I^{\text{0e}} \rangle+\langle E_I^{\text{1e}} \rangle +\langle E_I^{\text{2e}} \rangle\\
&=&  \langle E_I^{(1)}\rangle^2 
+ \sum_{p, q}^{\text{denom.}= 0} {\left|F_{pq}\right|^2 f_p^- f_q^+} 
\nonumber\\&&
+ \frac{1}{4}\sum_{p,q,r,s}^{\text{denom.}= 0} {\left|\langle pq || rs \rangle\right|^2f_p^- f_q^- f_r^+ f_s^+}. 
\label{E1E1_reduced}
\end{eqnarray}
Substituting this as well as Eq.\ (\ref{E2_reduced}) into Eq.\ (\ref{Omega_partial}), we arrive at the desired result [Eq.\ (\ref{Omega2_reduced})].


%
%

\end{document}